# Laser Induced Speckle as a Foundation for Physical Security and Optical Computing

Charis Mesaritakis
Dept. Information & Communication Systems Engineering
University of the Aegean
Karlovassi-Samos, Greece
cmesar@aegean.gr

Marialena Akriotou
Dept. Informatics & Telecommunications
National & Kapodistrian University of Athens, Athens, Greece
makriotou@di.uoa.gr

Dimitris Syvridis
Dept. Informatics & Telecommunications
National & Kapodistrian University of Athens, Athens, Greece
dsyvridi@di.uoa.gr

**We present a photonic system that exploits the speckle generated by the interaction of a laser source and a semi-transparent scattering medium, in our case a large-core optical fiber, as a physical root of trust for cryptographic applications, while the same configuration can act as a high-rate machine learning paradigm.**

## I. Introduction

The ever-expanding, highly heterogeneous and densely interconnected internet-of-things (IoT) ecosystem dictates for an equally expanding set of capabilities in terms of security and processing. Following this trend, a significant optimization effort has been invested during the last years, regarding the hardware of IoT components. In terms of security, the spotlight of attention has been focused on solutions like physical unclonable functions (PUFs) [1]. In a nutshell, PUFs are physical objects, whose complex physical structure, when properly probed, produces a response that is unique and cannot be used so as to evaluate the transfer function of the system. Therefore, they can be considered as the physical equivalents of mathematical one-way transformations.

The requirements that PUF modules should satisfy are: resiliency to physical cloning, meaning replication by an honest or malicious manufacturer, immunity against reverse engineering attacks, resiliency against modelling attempts, whereas their responses (multiple outputs from a single device) should be uncorrelated and thus satisfy strict criteria in terms of randomness [2]. Last but not least, a PUF module, contrary to conventional true random sources, should provide time-invariant responses, if the same pair of input and physical structure is used. Therefore, their most pronounced feature; is their ability to reproduce a response on demand, while this response retains all the qualities of a "true" random number. Based on these capabilities, PUFs have been used in applications ranging from cryptographic key generation [3], software-hardware interconnection [4], authentication tokens [5], and against code-reuse attacks [6].

The vast majority of PUFs are based on mature silicon technology due to footprint and cost considerations and their operation relies on fabrication related uncertainties in basic characteristics like gate-voltage, delay etc [7]. Despite their merits and compatibility with IoT devices, even state-of-the-art implementations are vulnerable to modelling and side channel attacks [7]. On the other hand, optical PUF's physical mechanism relies on the random speckle pattern generated when a laser beam illuminates a semi-transparent inhomogeneous material (fig.1) that is equipped with random defects (scatters) [1]. Security is preserved due to the complexity of the scattering process. Modelling attack scenarios demand solving Maxwell's equations for all possible physical arrangement [8]. Therefore, optical PUFs are considered vastly superior to their electronic counterparts and their security relies on "physics" and not on computationally techniques.

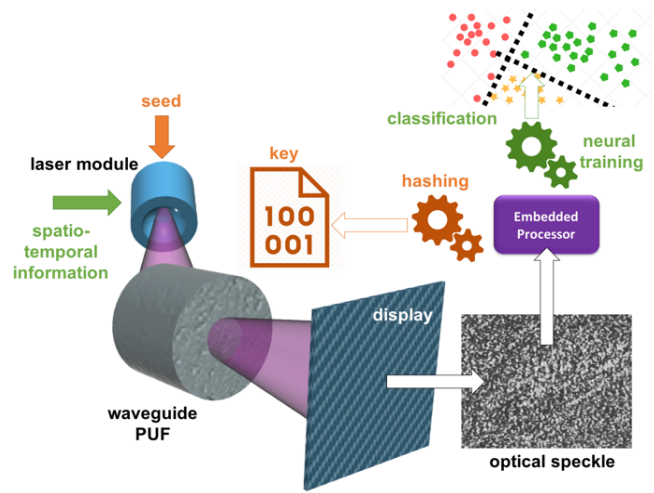

Figure 1. Basic schematic of a waveguide PUF utilized both as authentication key generator and spatio-temporal echo-state machine

Apart from physical layer security, a growing trend in the IoT landscape is the need for autonomous processing capabilities. This need contradicts the equally important demand for low-power, low-cost hardware. An interesting circumvention to this problem is the use of unconventional computational paradigms with disproportional efficiency compared to their available "hardware power", like: neural networks and neuromorphic systems. Moreover, the decentralized nature of IoT deployment suits really well to neural-like architectures where processing can be performed by remote nodes. Especially, the reservoir-computing (RC) paradigm is an interesting candidate due to the marginal requirements for training [9], the random nature of node's interconnections and the fact that it mainly targets the prediction/classification of time-evolving analogue and digital signals [10].

Thus it becomes evident that if a lightweight photonic system could be designed so as to perform both tasks, then merits like ultra-high bandwidth, high wall-plug efficiency and unparalleled physical layer security could be merged, thus strengthening the capabilities of IoT devices at the network-edge and consequently affecting the overall capabilities of the telecomm-infrastructure. In this context, we present results of a novel photonic configuration based on a large-core optical waveguide that can simultaneously play both roles, offering enhanced performance compared to conventional approaches.

## II. Waveguide Photonic Physical Unclonable Function

Recently, we demonstrated an optical PUF configuration based on a large-core optical waveguide [11]. The proposed scheme consisted of a commercial available step index polymer optical fiber with 980μm core and 20μm cladding. Both facets were processed through a noise-driven friction system, so as to exhibit random defects (fig.2a). The principle of operation involves the homogeneous illumination of the input facet by a single mode laser ($\lambda$=1540nm). Facet defects enable the random excitation of a significant number of transverse optical modes, whereas these modes propagate in the waveguide and interact through additional fabrication related defects (refractive index anomalies, cracks etc.). Transverse modes exiting the output facet exhibit difference phase velocity and random power distribution, thus their projection at an imaging device allows the recording of speckle-like random spatial pattern (modal-noise) (Fig.2b). The exact spatial profile of the speckle (response) is governed by the initial illumination conditions and by factors like the exact size-location of facet defects, in-fiber imperfections, bends, stress etc.

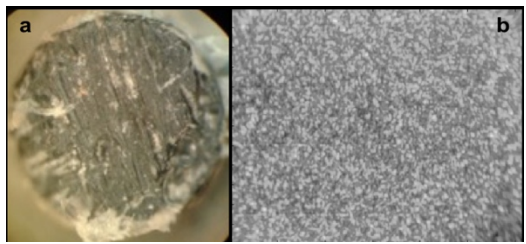

Figure. 2. a Photograph of the processed fiber's facet. b. image of a typical recorded optical speckle from the device.

### A. Authentication Applications

In [11] we evaluated the PUF's sensitivity to fabrication perturbations and therefore estimate the difficulty of physical replication. In particular, we fabricated 1000 PUFs and measured their responses under identical illumination conditions. We repeated the measurement without changing the illumination or the PUF specimen so as to evaluate the impact of noise. The raw PUF's responses (images) were processed through a typical hashing technique [8] and binary keys 256bit long were extracted for each image. In Fig.3 the fractional hamming distance of the first set (Fig.3 inter-class II) and the second (Fig.3 intra-class) are presented. It can be easily seen that the two distributions do not exhibit any overlap, thus the possibility for false-negative or false positives regarding the authenticity of a PUF's response is confirmed. Furthermore, it can be seen that mean value of the inter-class II is less than the optimum 0.5, resulting from experimental bias, possibly due to ambient light sources. The acquired results were compared with conventional optical PUFs, showing a reduction by 6 orders of magnitude in the probability of physical cloning; this performance improvement stems from the fact that the proposed PUF is a flexible waveguide and thus is extremely sensitive to in-fiber impairments like: refractive index variations, impurities, bends, stress etc.

### B. Deterministic Random Number Generation

The aforementioned results are very important for authentication applications, where the uniqueness of the physical structure guarantees security. On the other hand, by exploiting the deterministic response generation, PUFs can be also employed as alternative random number generators that alleviate the need for secure storage of the generated keys. Such an application offers an interesting alternative for various cryptographic protocols and cyber-shields systems (hacking attacks cannot lead to the extraction of critical cryptographic keys).

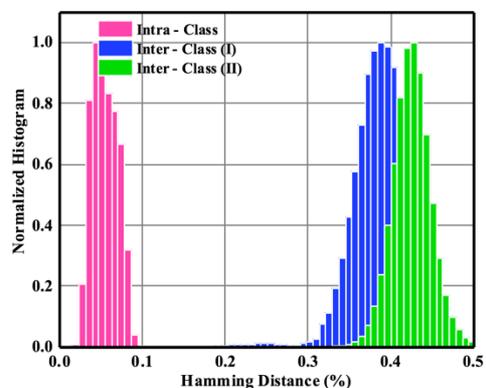

Figure 3. The computed fractional hamming distance for the generated keys by the waveguide PUF. Intra corresponds to bit-flips due to noise, inter-class I originate from the same specimen under different illumination conditions (wavelength tuning) and class-II correspond to keys obtained under identical laser bias for different PUF specimens.

Towards this direction, we replaced the single-mode laser source with a tunable laser having a tuning step of 10nm and wavelength range from 1540nm to 1570nm. Wavelength tuning has been employed as a challenge (input) generating mechanism, where each new wavelength varies the initial launching conditions and thus produce a new PUF response. In fig.3 (inter-class I) we present the fractional hamming distances computed for this dataset. It can be observed that their Gaussian distribution has a lower mean value ($\mu$=0.38) compared to class-II, due to the fact that the used step of $\Delta\lambda$=10nm allowed partially correlated responses. This fact can be easily amended by increasing the wavelength difference between consecutive responses. We evaluated the cryptographic quality of these keys using NIST statistical suit and typical information theory tools like minimum conditional entropy and have proven comparable to conventional pseudo-random sources (table I).

| Method | H-Conditional | H-minimum |
|---|---|---|
| Waveguide-PUF | 0.99 | 0.929 |
| SRAM-PUF | 1 | 0.937 |
| Ziggurat | 0.99 | 0.928 |

Table I: Conditional entropy and minimum conditional entropy for the waveguide-PUF, for the optimum silicon PUF and for a typical pseudorandom algorithm.

## III. Photonic PUF-RC Computational Paradigm

As mentioned above, the combination of cryptographic operations and actual processing would provide radical benefits to IoT devices. In particular, the RC machine-learning concept consists of a sub-category of multi-layer recursive neural networks (RNNs), where the hidden layer is equipped with a significant number of randomly interconnected nonlinear nodes and the output layer comprises a simple linear feedforward perceptron. The hidden layer forces the increase of the input's dimensionality, thus rendering classification or linear regression tasks straightforward and limit the training procedure only to the output layer [9]. A recent photonic implementation of an RC used simple waveguides and couplers, while the necessary node nonlinearity was introduced only through the square-law of the photo-detecting element [12].

Based on this implementation we propose a PUF-RC hybrid where the transverse optical modes are the operational equivalent of waveguides with random length and scattering centers act similar to optical couplers. Taking into consideration that the number of modes can be regulated during waveguide design and mode-coupling defects exhibit wavelength-size, the proposed PUF-RC hybrid scheme can offer enhanced scaling capabilities. The number of nodes (modes) can easily escalate to the $10^6$ while synaptic density can be drastically increased compared to planar RCs. Moreover, the inherent randomness of RCs fits remarkably well with the random nature of in-fiber defects.

In order to validate these claims, we have developed a travelling wave numerical model that incorporated the proposed PUF configuration in a short optical cavity so as to add memory effects. The model computed the spatial profile of all supported-modes and computed their evolution due to propagation and due to mode-coupling effects. As target inputs, we have generated video streams consisting of consecutive binary QR-code images. The generated time-evolving speckle patterns were fed to a trainable linear perceptron for classification. In order to generate a set of measurements for training we varied the signal-to-noise ratio (SNR) of the speckle patterns, assuming varying level of thermal noise at the photodiodes.

In fig.4 preliminary results regarding the classification error is computed versus the minimum SNR of samples used for different number of PUF defects. It can be seen that for zero defects classification error reaches a peak value of 75%, whereas for even a marginal number of mode-mixing defect (>15) the PUF-RC hybrid can classify the input video streams. Furthermore, noise has a detrimental effect on error, following an exponential decrease with SNR enhancement.

## IV. Conclusion

We present results concerning a photonic PUF scheme based on a multi-mode waveguide. The system offers lower probability of physical cloning compared to typical systems, and can be used as a deterministic random number generator. We also provide results that the system can be also used as an ultra-fast machine-learning paradigm able to classify time-evolving streams with minimum overhead.

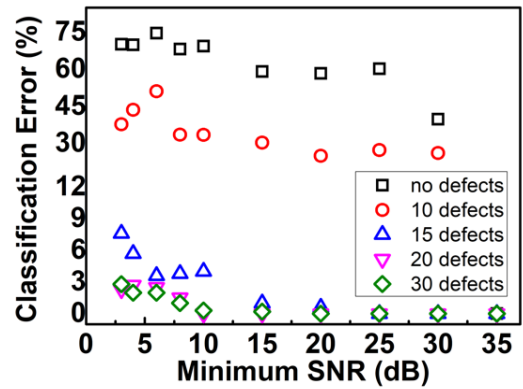

Figure 4. Classification error for binary video streams inserted to the PUF versus the minimum signal-to-noise ratio (SNR) of the samples used for neural training and for different number of mode-coupling enabling defects.

ACKNOWLEDGMENT

The research leading to these results has received funding from the European Union's Horizon 2020 research and innovation programme under grant agreement No 740931 (SMILE – Smart Mobility at the European Land Borders).